# High peak-power 2.1-µm femtosecond Holmium amplifier at 100 kHz

**Anna Suzuki,[†,*] Boldizsar Kassai,[†] Yicheng Wang, Alan Omar, Robin Löscher, Sergei Tomilov, Martin Hoffmann, Clara J. Saraceno**

*Photonics and Ultrafast Laser Science, Ruhr-Universität Bochum, Universitätsstrasse 150, 44801 Bochum, Germany*

*[†]These authors contributed equally to this work.*

*[*]anna.ono@ruhr-uni-bochum.de*

**High-power ultrafast laser sources in the short-wave infrared region are of great interest for numerous applications, including secondary sources of radiation and processing of materials commonly opaque in the near-infrared region. In this wavelength region, direct laser amplification around 2.1-µm wavelength within the atmospheric transparency window is particularly attractive for realizing compact and efficient high-power lasers. However, this wavelength region was widely underrepresented in femtosecond laser technology so far. Here, we report on a 2.1-µm laser system delivering 97-fs pulses with an unprecedented combination of high peak power of 525 MW and high repetition rate of 100 kHz. The amplifier system consists of a mode-locked oscillator seeding a regenerative amplifier (RA) using the novel broadband material Holmium (Ho)-doped $CaAlGdO_4$ (CALGO), operating in the chirped pulse amplification (CPA) scheme and a nonlinear compression stage based on a Herriott-type multi-pass cell (MPC) with bulk material. We demonstrate the potential of this unique laser system by generating a micro plasma in ambient air, demonstrating its high intensity for future plasma-driven secondary sources. This system bridges the gap between conventional 1-to-10 kHz amplifiers and high-power MHz laser oscillators in this attractive wavelength range.**

Ultrafast 2-µm lasers with high peak powers and high repetition rates are attractive driving sources for nonlinear wavelength conversion, e.g. efficient mid-infrared generation using non-oxide nonlinear crystals [1], high-harmonic generation toward higher photon energy regions [2], and terahertz generation based on two-color plasma filaments [3]. They are also in high demand for material processing applications, in particular for materials that are only transparent at long wavelengths, such as silicon [4]. Conventional sources used for these applications in this spectral region are mostly based on optical parametric amplifiers (OPA) pumped by near-infrared Ti:sapphire or Yb-based lasers [5–7], which are inefficient and often limited in repetition rate, and usually suffer from complicated space-time couplings that degrade intensity. Direct laser emission and amplification in this wavelength region is an attractive alternative for efficiency, simplicity of the whole system, as well as efficient repetition rate scaling. Most efforts so far in this area have focused on diode-pumped Tm-doped fiber amplifier architectures. A femtosecond fiber CPA demonstrated 1.65-mJ pulse energy and 167-W average power at a 101-kHz repetition rate [8], and 1-kW average power was achieved at a repetition rate of 80 MHz [9]. However, the operation wavelength of the Tm-doped fiber amplifier is around 1.95 µm where strong water vapor absorption exists, resulting in significant beam degradation in both phase and spatial beam quality at high power levels with free-space propagation [10]. Ho-doped gain materials are, in this regard, very promising for high-power, high-energy amplifier systems due to their operation wavelength of 2.1 µm. This is attractive for applications as this coincides with an atmospheric transmission window, facilitating power and energy scaling, as well as beam transport. In addition, they exhibit relatively high gain cross sections and long upper-level lifetimes of several ms, indicating their huge energy storage capability. Moreover, they can be pumped by high-power 1.9-µm Tm fiber lasers that are widely available, and their small quantum defect of less than 10% leads to a small thermal load and highly efficient laser operation, making them suitable for high-average power lasers.

So far, 2.1-µm Ho laser amplifiers have been demonstrated utilizing Ho:YLF or Ho:YAG materials in the kHz repetition rate range. Figure 1 shows the overview of the peak power of Ho-based amplifiers at different repetition rates. Ho:YLF in particular is a gain material with a long upper-level lifetime of 14 ms, a small nonlinear refractive index $n_2$, and a negative and small thermo-optic coefficient [11,12], making it a workhorse in most amplifier systems at this wavelength. A Ho:YLF RA in CPA configuration seeded by a three-stage OPA pumped by a Yb:KGW CPA demonstrated 22.5-mJ, 2-ps pulse amplification at a center wavelength of 2050 nm at 1 kHz, reaching a peak power of 10 GW [13]. 34-GW, 75-mJ pulse amplification with 2.2-ps pulse duration was achieved with a RA followed by two booster amplifiers at 1 kHz [14]. A CPA and a cryogenically-cooled booster amplifier delivered 220-mJ pulse energy with 16-ps pulse duration at a low repetition rate of 100 Hz [15]. With a CPA-free arrangement and 5-ps pulse seeding, a multi-pass amplifier with a dual-crystal booster amplifier generated pulse energy of 1.2 mJ, 55 µJ, and 104 µJ with pulse duration of 8.3 ps, 8.2 ps, and 6.4 ps, at 1 kHz, 10 kHz, and 100 kHz repetition rates, respectively [16].

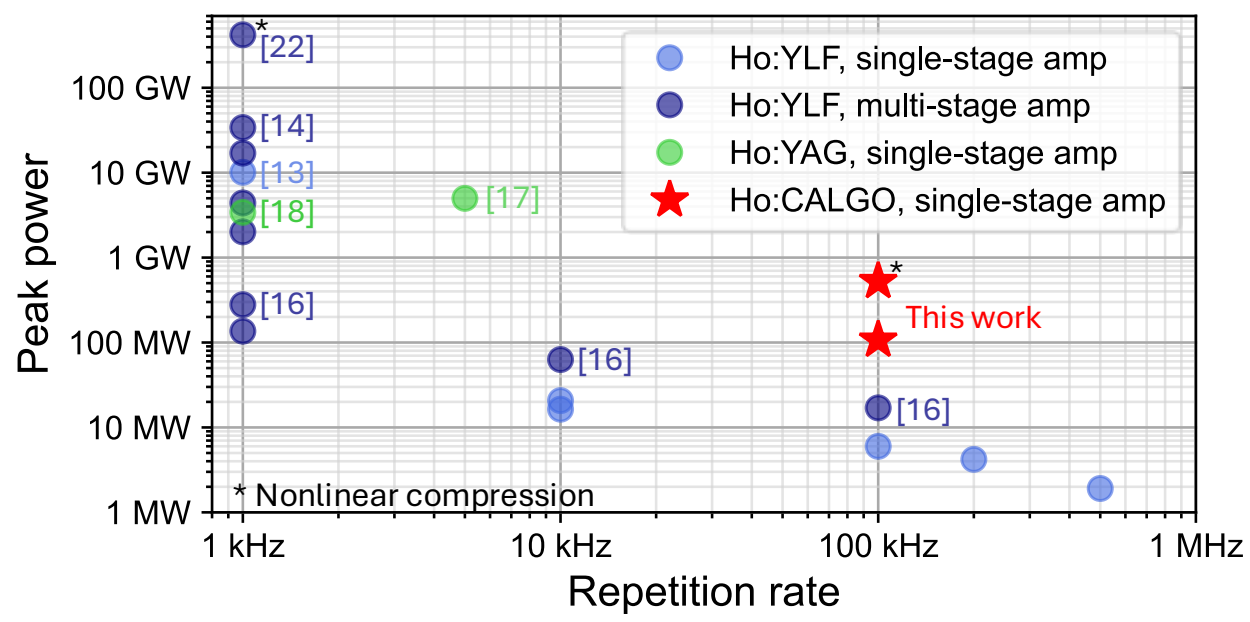

Fig. 1. Overview of Ho-based amplifier systems.

Despite all these excellent results, the main challenge in Ho-based amplifier systems is the available pulse duration, limited to the multi-ps range as shown above. For applications seeking short pulse durations, multiple complex compression stages are required to use such laser systems. This explains why so far most of these systems are used for pumping further OPCPAs, instead of for direct applications, leaving a huge untapped potential. The narrow and structured gain spectra of Ho-doped materials in traditionally used hosts lead to strong gain narrowing, especially in high-gain amplifiers. Sub-ps Ho-based RAs were only realized using a Ho:YAG crystal which exhibits a rather narrow and structured gain profile in comparison to Ho:YLF. By using additional spectral and phase pre-shaping by means of the Dazzler, multi-GW peak power level, 530-fs, 1-mJ pulses at 5 kHz [17] and 980-fs, 3.8-mJ pulses at 1 kHz were obtained [18]. However, using the Dazzler, one sacrifices significant seed energy thus requiring additional pre-amplifier stages.

In this regard, the newly demonstrated Ho:CALGO crystal is an attractive alternative for ultrafast laser amplifiers. It exhibits a broad and flat gain profile due to inhomogeneous spectral broadening caused by a disordered crystal structure [19], which is favorable for broadband pulse amplification. Therefore, sub-ps pulse amplification is expected without special spectral shaping techniques. To date, only mode-locked laser oscillators using Ho:CALGO have been demonstrated, achieving the shortest pulse duration with watt-level average output powers [20,21], indicating the potential of the gain medium.

In this work, we demonstrate a high-power broadband RA using a Ho:CALGO crystal at 2.1 µm. Direct seeding with a 2.1-µm mode-locked oscillator and a conventional CPA arrangement enables a simple laser amplifier delivering an average output power of 11.2 W at a repetition rate of 100 kHz, corresponding to a pulse energy of 112 µJ. Although this RA has a large gain factor of $10^{5.4}$, the broad gain bandwidth of Ho:CALGO enables sub-ps pulse amplification without any spectral shaping or preamplification.

The short pulse duration from the RA directly allows us to further implement simple pulse-shortening techniques for peak power enhancement, enabling applications requiring gas/air ionization and or shorter pulses in general. Previous results of nonlinear pulse compression at this wavelength focused on high energy schemes based on gas-filled hollow-core fiber and succeeded in compressing multi-10 mJ, 3-ps pulses at 1 kHz down to 86 fs, however, it showed a relatively low transmission of 58.7% [22]. In this regard, MPCs are an attractive technique which provides high optical transmission and excellent preservation of the beam quality, and has so far been demonstrated using a gas-filled MPC with high-power Tm fiber laser amplifier systems [23,24]. We implement a nonlinear pulse compression technique utilizing a Herriott-type bulk MPC for the first time at this wavelength. Using a YAG plate as a nonlinear medium, a compressed pulse duration of 97 fs was obtained with excellent optical transmission of 90%, and the peak power reached 525 MW, demonstrating the first bulk MPC for high-power 2-µm lasers. In view of future applications of this system, we demonstrate sufficient focused peak intensity to generate a plasma in the air using a reflective microscope objective, demonstrating the potential of this system for a plethora of future experiments making use of fs-laser-driven plasmas, for example, efficient THz generation driven at long wavelengths [25].

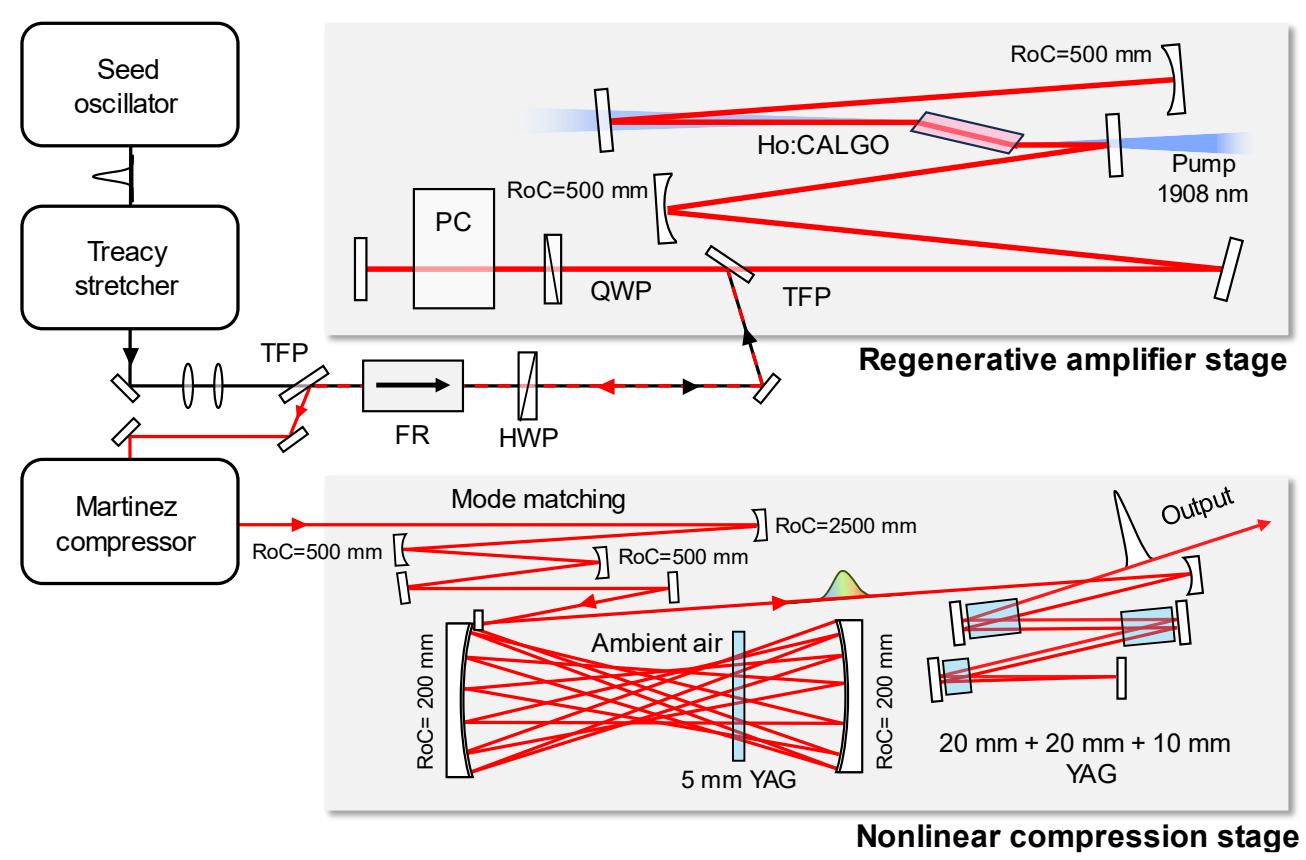

Fig. 2. Schematic of the Ho:CALGO RA and nonlinear pulse compression stage based on a bulk MPC.

The amplifier system including the nonlinear compression stage is shown in Fig. 2. As a seed source, we used a home-built SESAM mode-locked Tm,Ho:CLNGG laser oscillator [26], which delivers pulses as short as ≈280 fs with an average output power of 95 mW at a repetition rate of 70.3 MHz, corresponding to a pulse energy of 1.35 nJ. The seed pulses were negatively stretched by a grating pair with a line density of 600 lines/mm in the Treacy configuration. It provides a group delay dispersion (GDD) of about -22 $ps^2$, resulting in a stretched pulse duration of ≈200 ps. After the stretcher, the seed pulse had a reduced pulse energy of 0.48 nJ, and the center wavelength and spectral bandwidth were 2093 nm and 16 nm for full-width at half maximum (FWHM), respectively. The seed pulse was coupled into the RA cavity through a thin-film polarizer (TFP), a Faraday rotator (FR), and a half-wave plate (HWP). The amplifier stage consists of a linear cavity including a Pockels cell (PC) (bpp2.5c3, Bergmann Messgeräte Entwicklung KG) employing an RTP crystal (Leysop Ltd.) and a quarter-wave plate (QWP) for regenerative amplification. The RA cavity length is ≈2.03 m corresponding to a repetition rate of ≈74 MHz, which is slightly greater than that of the seed oscillator, so that we can pick up and confine only one pulse in the RA cavity without additional pulse picking. As a gain material, we used a Brewster cut 1-at.% $Ho^{3+}$-doped CALGO crystal with a dimension of 3×3×27.1 $mm^3$, used with π-polarization geometry (E||c). The pump source is a continuous-wave single-mode Tm fiber laser (IPG Photonics) operating at 1908 nm, which was the only one available at the time of the experiments. Note that this wavelength is not the peak absorption wavelength of 1950 nm for Ho:CALGO [19], offering a straightforward path for future improvement of the efficiencies reached here. The pump absorption amounted to ≈43% at this wavelength. The beam spot size in the gain crystal was 145×270 $\mu m^2$ in radius for both pump and laser. The amplified pulses were extracted from a TFP and subsequently linearly compressed by a Martinez compressor composed of a grating pair which is the same as the initial stretcher and concave mirrors with a radius of curvature (RoC) of 750 mm. Note that for the first experimental realization, we built a Treacy stretcher for simplicity and flexibility in the alignment, which is uncommon in conventional CPA systems. However, the sign of the pre-chirping does not have a large influence here. In fact, the amount of GDD given by the stretcher is significantly larger than that of the RA cavity (-0.31 $ps^2$ for 28 round trips, RTs), therefore the cavity GDD does not change the pulse duration significantly during the amplification.

For the nonlinear compressor stage, mode-matching optics were placed in front of the MPC which is composed of three concave mirrors. The Herriott-type MPC consisted of two 2-inch concave mirrors with RoC of 200 mm and a 5-mm anti-reflection coated YAG plate to achieve spectral broadening via self-phase modulation (SPM). YAG was chosen as a compromise between sufficiently high $n_2$, a moderate GDD at 2.1 µm, and good optical quality among commercially available materials. Two concave mirrors are placed with a separation of 375 mm, enabling 13 round trips (RT) of the input pulses. After the MPC, the temporal compression stage was placed to compensate for residual chirp using the material dispersion. Three YAG plates (two 20-mm and one 10-mm) with 4-passes provide a total GDD of -15500 $fs^2$ at 2080 nm.

The RA experiments were performed at a 100-kHz repetition rate. The number of RTs was 28, which is optimized by a simulation using the model in Ref. [27], and experimentally confirmed (see supplemental document). The average power and pulse energy evolution depending on the incident pump power are shown in Fig. 3(a). The simulation and the experimental results show good agreement with each other. At an incident pump power of 71 W, a maximum average output power of 11.2 W was obtained, corresponding to a pulse energy of 112 µJ (Fig. 3(a)). The B-integral was estimated to be 0.48. Further power scaling was limited by the risk of damage due to the thermal load in the gain crystal. Other gain geometries which offer superior cooling efficiency such as disk or slab would solve this issue. Considering the pump absorption efficiency of ≈43%, the optical-to-optical conversion efficiency was about 36%. This relatively low efficiency compared to the Stokes limit is most likely due to the quenching effect caused by energy transfer upconversion [28]. To increase the laser efficiency, a lower doped gain crystal would be beneficial. This however imposes the use of longer crystals, resulting in other design constraints and limitations to be weighed in, such as increased B-integral. The amplified pulse spectrum is shown in Fig. 3(b). Compared to the seed spectrum, the center wavelength was blue-shifted according to the gain profile of Ho:CALGO and the spectrum got slightly narrower to about 11 nm for FWHM. Although the amplifier shows a high gain factor of $10^{5.4}$, gain narrowing was not prominent and the broadband spectrum supported the Fourier transform limited (FTL) pulse duration of 626 fs.

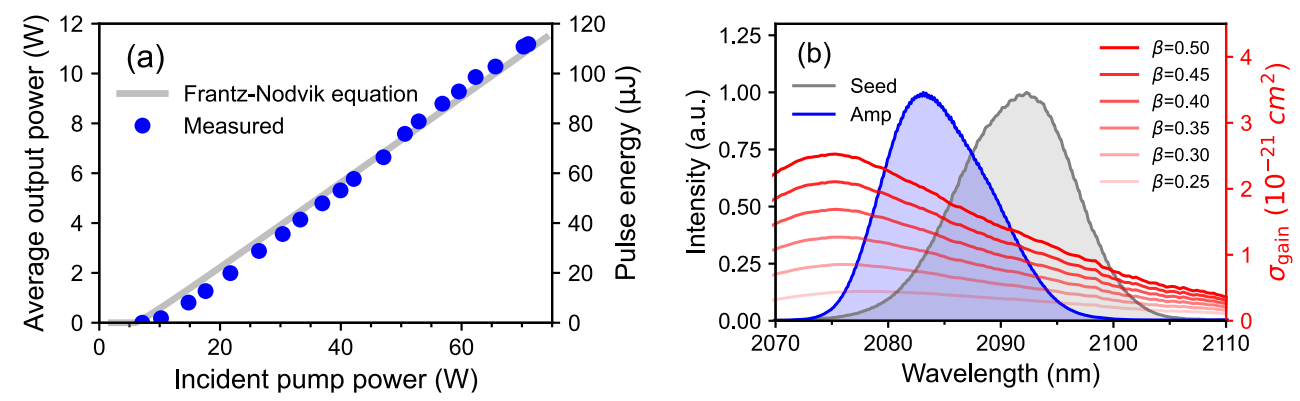

Fig. 3. (a) Simulated (grey) and measured (blue) average power and pulse energies of the Ho:CALGO RA dependent on the incident pump power at 100 kHz. (b) Seed (grey) and amplified (blue) pulse spectra, and gain spectra of Ho:CALGO for π-polarization (red).

The amplified pulses were compressed by the Martinez compressor. The energy throughput of the compressor was 83%, resulting in a pulse energy of 93 µJ after the compressor. Detailed pulse characterization using the second-harmonic frequency-resolved optical gating (FROG) technique, along with output stability and beam quality, is provided in the supplemental document. The retrieved pulse duration was 750 fs for FWHM and the peak power reached 107 MW.

For the nonlinear compression experiments, we limited the incident power to 8 W to avoid conical emission at the YAG plate at the end of the temporal compression stage, which would otherwise have led to beam degradation. This could not be mitigated by increasing the beam size due to the limited aperture of our 1-inch optics. The nonlinear compression stage showed an overall high transmission of 90%, resulting in 7.2-W average power and 72-µJ pulse energy. The nonlinearly broadened spectrum spans from 2000 nm to 2150 nm after 13 RTs in the MPC. Subsequently, the temporal compression was done by utilizing YAG plates. The results of the nonlinear pulse compression are shown in Fig. 4. The measured and retrieved FROG traces show good agreement as shown in Fig. 4(a, b). The retrieved temporal pulse profile was evaluated to a pulse duration of 97 fs, which is close to the FTL pulse duration of 91 fs. The calculated peak power reached 525 MW, corresponding to a peak power enhancement factor of 4.9. The $M^2$ value was measured to be less than 1.1 for both axes (see supplemental document). It is noteworthy that we opted for a moderate compression factor in this study, to efficiently confine the energy within the main pulse. Although we could further enhance the spectral broadening by moving the YAG plate in the MPC close to the beam focus, this did not increase peak power and pulse shortening was only possible with significant pedestals. Therefore, we opted for optimal peak power and pulse shape at more moderate compression ratios.

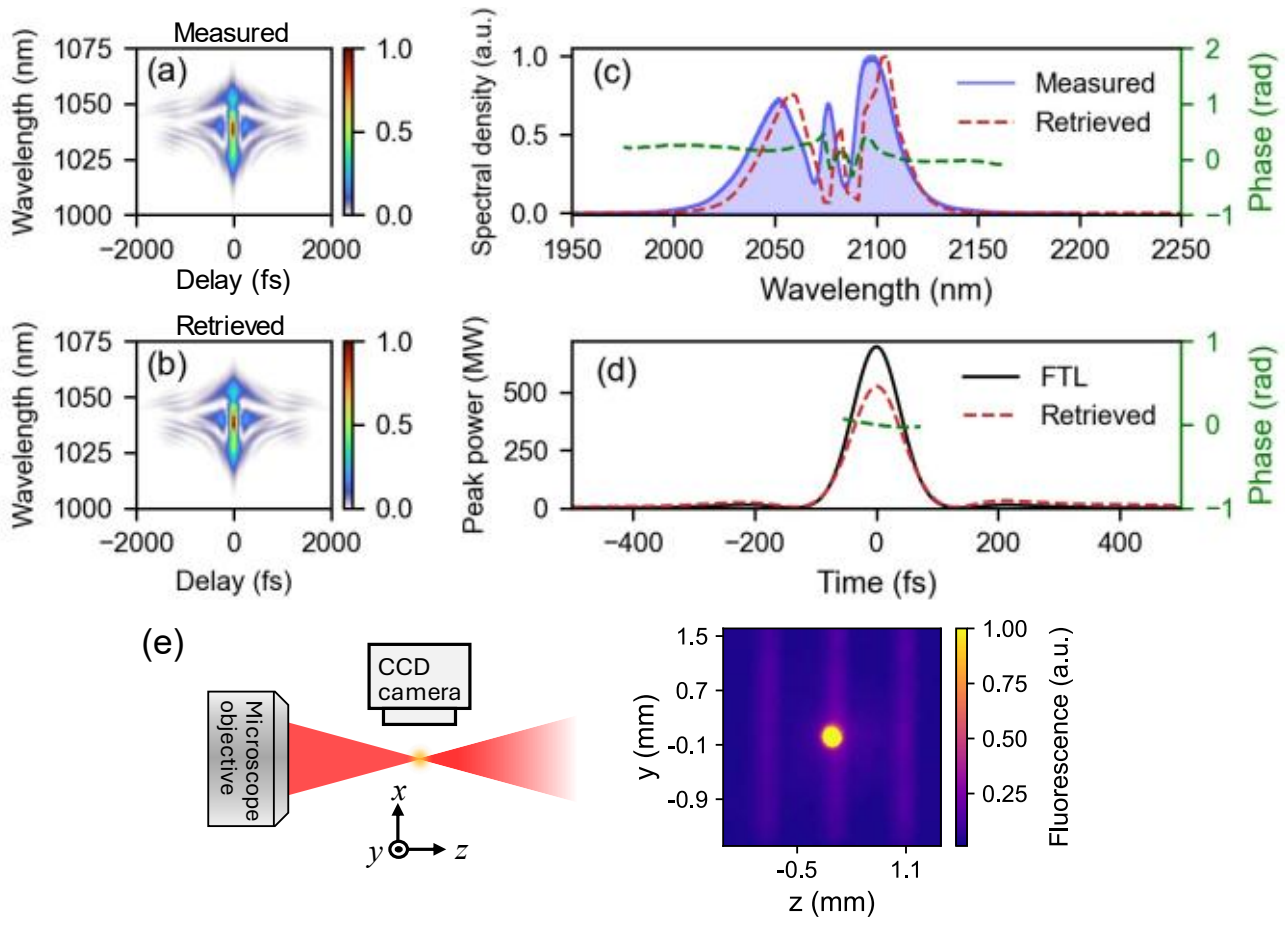

Fig. 4. Characterization of the compressed pulses of the Ho:CALGO RA after the nonlinear compressor. (a) Measured and (b) retrieved FROG traces on a grid of 512×512 and an error of 0.54%. (c) Measured (blue) and retrieved (red) spectra, and (d) retrieved temporal profile (red) and calculated FTL pulse (black). (e) Experimental setup for plasma generation (left) and CCD image of plasma fluorescence generated by 72-µJ, 97-fs pulses at 100 kHz with a background of ruler lines (right).

Finally, we demonstrate the formation of a micro plasma as a preliminary experiment for future applications in THz generation. We focused the beam using a reflective microscope objective with a focal length of 8 mm and NA of 0.4. Using 72-µJ, sub-100-fs pulses, plasma fluorescence was visibly observed at the working distance of the objective, and the fluorescence image was captured using a CCD camera as shown in Fig. 4(e). As secondary verification, we measured the ionic charge collected by a high-field bias using a capacitive plasma probe [29]. The probe electrodes were placed around the micro plasma, and the current proportional to the ionic charge was measured to be (666.08±90.49) nA, which was an order of magnitude higher than the measurement noise of (76.02±1.44) nA, confirming ionization of the ambient air and the high peak intensity of our laser system for future applications.

In summary, we demonstrate 11.2-W, 112-µJ broadband RA at 100 kHz based on Ho:CALGO and a basic CPA arrangement. Thanks to the broad and flat gain profile of Ho:CALGO due to inhomogeneous broadening, 750-fs pulse amplification was achieved without any spectral shaping technique, which allows us to reach a 107-MW peak power directly from the laser system. This result represents the highest average power and highest peak power in Ho-based amplifiers at a 100-kHz repetition rate. Furthermore, we demonstrate the nonlinear pulse compression using a bulk MPC for the first time in the 2-µm wavelength range. After the nonlinear compression stage, pulses as short as 97-fs were obtained with an excellent optical transmission of 90% and good beam quality with $M^2$ <1.1. It enabled the peak power enhancement to 525 MW, even allowing for ionization of ambient air. This all-bulk amplifier and MPC configuration are realized in a simple and robust system, making it a highly attractive high-peak-power, high-repetition-rate, 2.1-µm source ideally suited for high-power secondary sources of radiation, for example, efficient THz sources with ultra-broad bandwidths and high efficiency.

**Funding.** Ruhr-Universität Bochum (Open Access Publication Funds); Deutsche Forschungsgemeinschaft (390677874); European Research Council (ERC) (101138967- Giga2u).

**Acknowledgment.** We would like to thank Dr. Ignas Stasevičius, Dr. Ignas Astrauskas, and Dr. Julius Darginavičius from Light Conversion for fruitful discussions.

**Disclosures**. The authors declare no conflicts of interest.

**Data Availability Statement (DAS).** Data underlying the results presented in this paper is available in Ref. [30].

**Supplemental document.** See Supplement 1 for supporting content.

## Full references

## High peak-power 2.1 μm femtosecond Holmium amplifier at 100 kHz: supplemental document

This supplemental document addresses details about the experimental results.

### 1. Simulation and experiments for finding the optimum number of round trips

The optimum number of RTs was estimated by a simulation based on the spectrally resolved Frantz-Nodvik equation [1], and experimentally confirmed by changing the number of RTs, which revealed an optimum number of RTs of 28 for maximum energy extraction in this case [(Fig. S1(a))]. The bifurcation instability was not observed in the experiment, and the onset of the bifurcation was predicted at higher RTs of 81 at this repetition rate as shown in Fig. S1(b). This is because the 100 kHz repetition rate is much larger than the inverse of the fluorescence lifetime of Ho:CALGO (≈6 ms) [2,3]. Therefore, the efficient energy extraction was not disturbed by the bifurcation instability at this repetition rate.

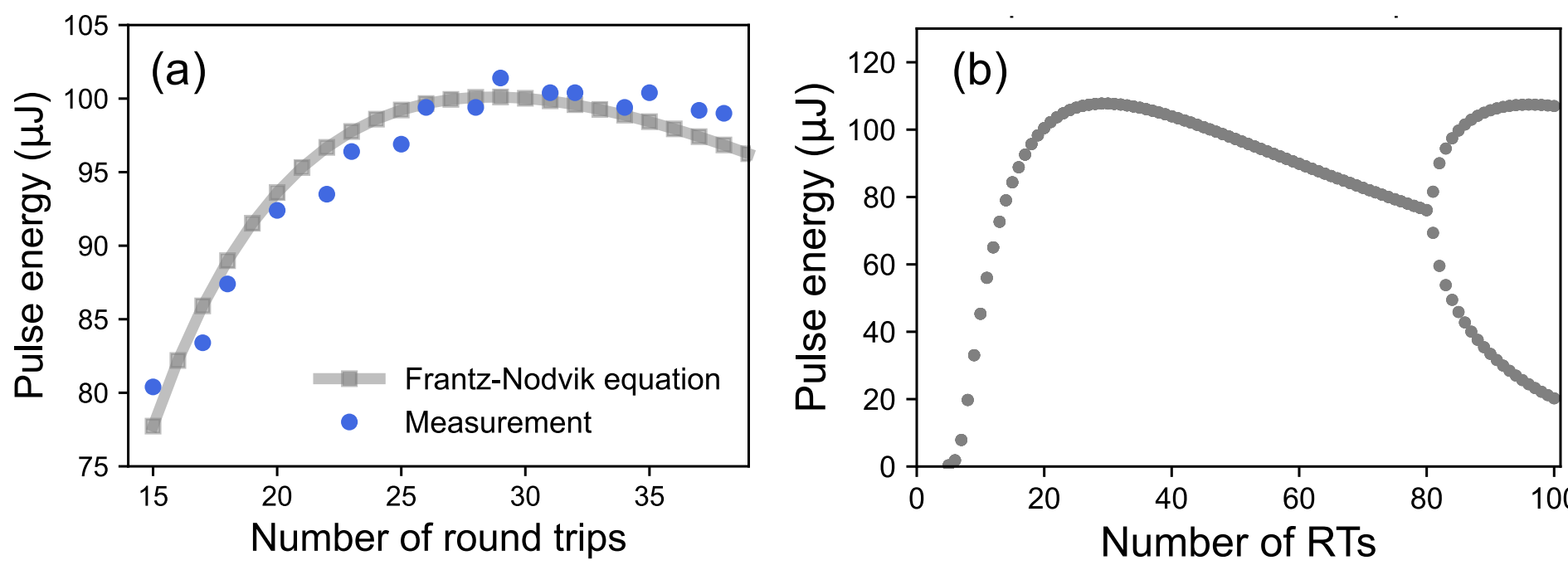

Fig S1 .(a) Simulated (grey) and measured (blue) pulse energies of the Ho:CALGO RA dependent on the number of RTs. (b) Simulated bifurcation diagram at 100 kHz.

### 2. FROG measurement of the regenerative amplifier output

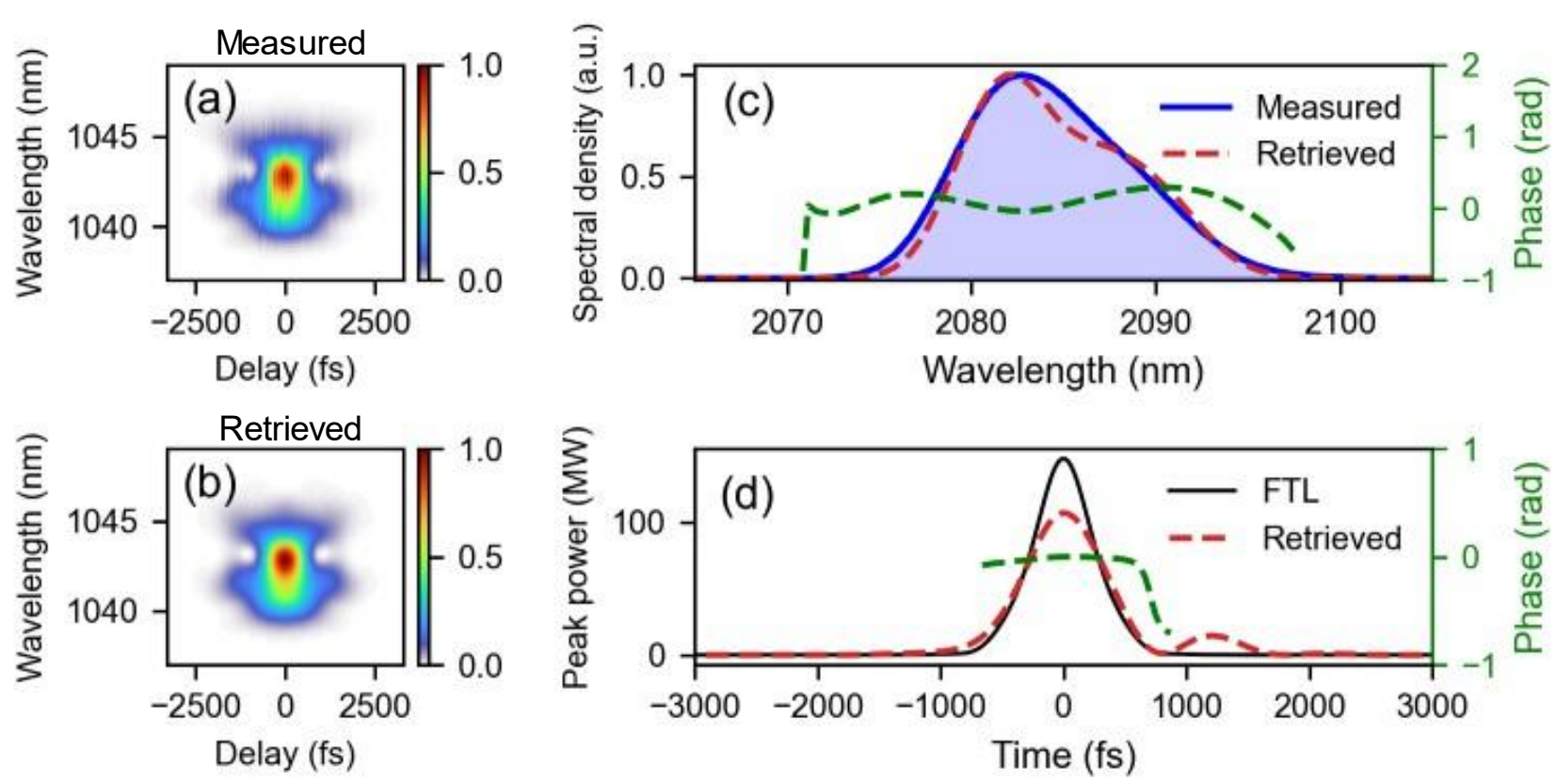

Fig. S2 Characterization of the Ho:CALGO RA output after the Martinez compressor. (a) Measured and (b) retrieved FROG traces. (c) Measured (blue) and retrieved (red) spectra, and (d) retrieved temporal profile (red) and calculated FTL pulse (black).

The output pulses after the Martinez compressor were characterized using a home-built second-harmonic frequency-resolved optical gating (FROG) setup as shown in Fig. S2. The retrieved FROG trace on a grid size of 512×512 exhibits an error of 0.25%. The measured and retrieved spectrum show good agreement as shown in Fig. 2(c). The retrieved temporal pulse exhibits a pulse duration of 750 fs for FWHM with a small sub-pulse indicating a residual cubic spectral phase. The calculated peak power reaches 107 MW.

## 3. Output power stability of the regenerative amplifier

To evaluate the stability of the RA output, the pulse energy was recorded for 30 minutes as shown in Fig S2. The root mean square (RMS) fluctuations were below 0.39% for both the output directly from the RA and the output from the Martinez compressor. Although the whole system was operating in ambient air, it showed good stability.

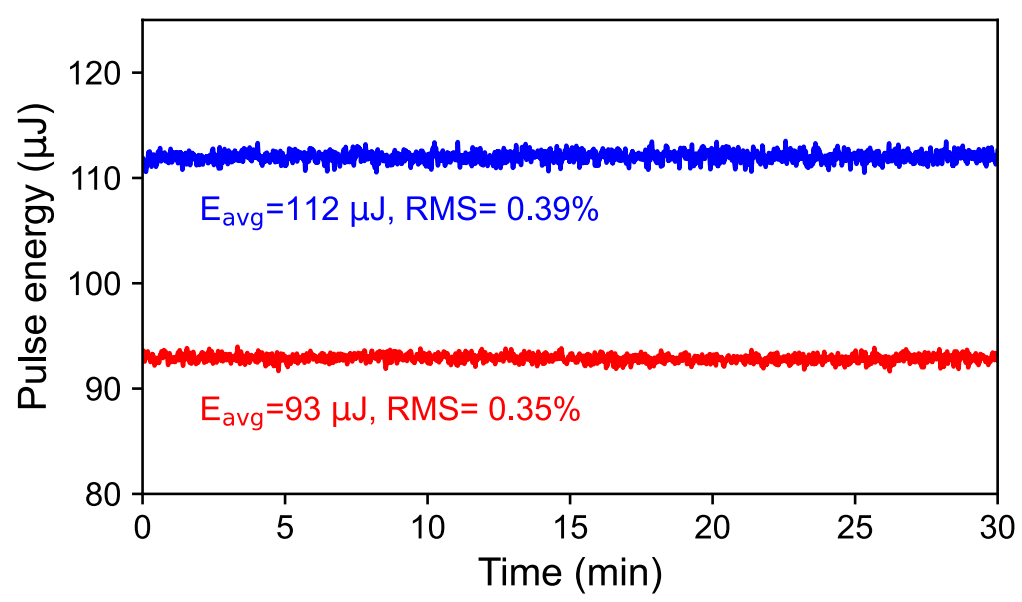

Fig. S3. Long-term output stability of the RA (red) and after the Martinez compressor (blue)

## 4. Beam quality $M^2$ measurement of the regenerative amplifier and the multi-pass cell output

The beam quality of the amplifier output was measured according to ISO 11146 standard with a microbolometer camera (WinCamD, Dataray). Figure S3 shows beam radii of the RA and MPC output. From the fitting curves, the $M^2$ values of the RA output were determined to be 1.19×1.11 for the horizontal and vertical axis, indicating a beam quality close to the $TEM_{00}$ mode. The $M^2$ values of the MPC output were 1.07×1.09, indicating that the MPC even cleaned up the spatial beam quality.

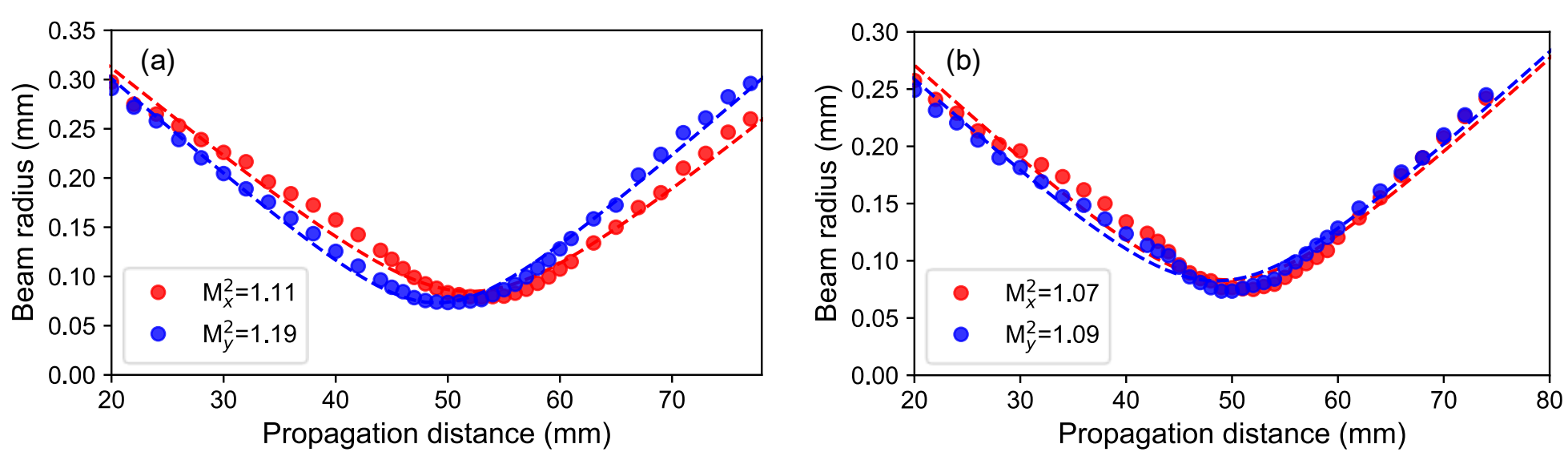

Fig. S4. Beam propagation of RA output after the Martinez compressor (a) and the MPC.